\documentclass[sigconf]{acmart}

\renewcommand\footnotetextcopyrightpermission[1]{}
\acmConference[]{}{}

\usepackage{multirow}
\usepackage{graphicx}
\usepackage[skip=10pt]{caption}
\usepackage[nohyperlinks, nolist]{acronym}
\usepackage{tabularx, booktabs}

\usepackage[english]{babel}
\usepackage[utf8]{inputenc}

\usepackage{tikz}
\usetikzlibrary{positioning}

\newcolumntype{Y}{>{\centering\arraybackslash}X}

\begin{acronym}
\acro{supi}[SUPI]{Subscription Permanent Identifier}
\acro{suci}[SUCI]{Subscription Concealed Identifier}
\acro{guti}[GUTI]{Global Unique Temporary Identifier}
\acro{ran}[RAN]{Radio Access Network}
\acro{mac}[MAC]{Message Authentication Code}
\acro{rrc}[RRC]{Radio Resource Control}
\acro{aka}[AKA]{Authentication and Key Agreement}
\acro{gnb}[gNB]{gNodeB}
\acro{ue}[UE]{User Equipment}
\acro{mme}[MME]{Mobility Management Entity}
\acro{as}[AS]{Access Stratum}
\acro{nas}[NAS]{Non-Access Stratum}
\acro{ims}[IMS]{IP Multimedia Subsystem}
\acro{iot}[IoT]{Internet of Things}
\acro{sim}[SIM]{Subscriber Identity Module}
\acro{lmf}[LFM]{Location Management Function}
\acro{sba}[SBA]{Service-Based Architecture}
\acro{sbi}[SBI]{Service-Based Interface}
\acro{amf}[AMF]{Access and Mobility Function}
\acro{upf}[UPF]{User-Plane Function}
\acro{nef}[NEF]{Network Exposure Function}
\acro{sepp}[SEPP]{Secure Edge Protection Proxy}
\acro{dn}[DN]{Data Network}
\acro{nrf}[NRF]{Network Repository Function}
\acro{smf}[SMF]{Session Management Function}
\acro{pfcp}[PFCP]{Packet Forward Control Protocol}
\acro{plmn}[PLMN]{Public Land Mobile Network}
\acro{udm}[UDM]{Unified Data Management}
\acro{pws}[PWS]{Public Warning System}
\acro{sib}[SIB]{System Information Broadcast}
\acro{pei}[PEI]{Permanent Equipment Identifier}
\acro{smc}[SMC]{Security Mode Command}
\acro{tls}[TLS]{Transport Layer Security}
\acro{ip}[IP]{Internet Protocol}
\acro{ftp}[FTP]{File Transfer Protocol}
\acro{pdcp}[PDCP]{Packet Data Convergence Protocol}
\acro{gtp-c}[GTP-C]{Generic Tunnel Protocol-Control Plane}
\acro{http}[HTTP]{Hyper Text Transfer Protocol}
\acro{udp}[UDP]{User Datagram Protocol}
\acro{pcf}[PCF]{Policy Control Function}
\acro{ausf}[AUSF]{Authentication Server Function}
\acro{af}[AF]{Application Function}
\acro{nf}[NF]{Network Function}
\acro{uicc}[UICC]{Universal Integrated Circuit Card}
\acro{aes}[AES]{Advanced Encryption Standard}
\acro{pb3c}[PB3C]{Piggy-Backing-based Command and Control Communication}
\acro{im3c}[IM3C]{Induced Messages-based Command and Control Communication}
\acro{5gpp}[5GPP]{5G Puppeteer Protocol}
\acro{seaf}[SEAF]{Security Anchor Function}
\acro{ngap}[NGAP]{Next Generation Application Protocol}
\acro{aesni}[AES-NI]{Advanced Encryption Standard New Instructions}
\end{acronym}

\usepackage{tikz}
\usepackage{tikzpeople}
\usetikzlibrary{decorations.text}
\usetikzlibrary{shapes}
\usetikzlibrary{decorations.pathreplacing,calc}

\begin{document}

\title{5G Puppeteer: Chaining Hidden Command and Control Channels in 5G Core Networks}

\author{Julian Sturm}
\email{Julian.Sturm@zitis.bund.de}
\affiliation{
    \institution{ZITiS}
    \city{Munich}
    \country{Germany}
}

\author{Daniel Fraunholz}
\email{Daniel.Fraunholz@zitis.bund.de}
\affiliation{
    \institution{ZITiS}
    \city{Munich}
    \country{Germany}
}

\author{Oliver Zeidler}
\email{Oliver.Zeidler@tum.de}
\affiliation{
    \institution{Technical University of Munich}
    \city{Munich}
    \country{Germany}
}

\author{Katharina Schaar}
\email{Katharina.Schaar@zitis.bund.de}
\affiliation{
    \institution{ZITiS}
    \city{Munich}
    \country{Germany}
}

\author{Wolfgang Kellerer}
\email{Wolfgang.Kellerer@tum.de}
\affiliation{
    \institution{Technical University of Munich}
    \city{Munich}
    \country{Germany}
}

\begin{abstract}
    Mobile networks are essential for modern societies.
    The most recent generation of mobile networks will be even more ubiquitous than previous ones.
    Therefore,     the security of these networks as part of the critical infrastructure with essential communication services is of the uttermost importance.
    However,     these systems are still vulnerable to being compromised, as showcased in the recent discussion on supply chain security and other challenges.
    This work addresses problems arising from compromised 5G core network components.
    The investigations reveal how attacks based on command and control communication can be designed so that they cannot be detected or prevented.
    This way, various attacks against the security and privacy of subscribers can be performed for which no effective countermeasures are available.

\end{abstract}



\maketitle

\small
\section{Introduction}
\label{intro}

5G is an essential enabler for modern telecommunication services.
The technology will be used for all types of communication from smartphones and smartwatches to smart homes and smart cities.
The \ac{iot} will rely heavily on cellular communication.

Due to its vital role, the 5G infrastructure is considered critical infrastructure and requires careful protection.
The severe negative implications of malfunctioning networks make them high value targets for attackers.
Additionally, a high level of security is difficult to achieve due to the high complexity of the 5G infrastructure:
There are legacy technologies like 2G interconnected, there are foreign networks connected via roaming, there are hard- and software supply chains with potential security risks, there is a large, geographically spread physical infrastructure, and there are all other security threats associated to computer systems, like social engineering and software vulnerabilities.
Therefore, besides the prevention of attacks the detection of compromised components and the mitigation of their impact are important.
This is particularly true as a 5G network might be compromised a long time before the actual attacks, i.e., previously installed backdoors might be activated during political tensions or wartime.

The command and control of compromised network structures is a long-term research topic with much previous research \cite{WangDetectioncommandcontrol2016, FedynyshynDetectionClassificationDifferent2011}.
However, 5G-specific command and control has not been researched yet although being much more dangerous as it cannot be easily distinguished from genuine 5G communication and, thus, is hard to prevent and detect.
In this work, we propose an approach to provide command and control capabilities in a (partly) compromised 5G core network to execute different attacks in the network at any time required.
Exemplary attacks are the extraction of keys to decrypt communication, the abuse of the \ac{pws} to send false alert messages to the public, or the obtaining of the geolocation of users to track them.

The research questions addressed are:
1. Is it possible to establish command and control communication into and within a 5G core network by using standardized or genuine 5G interfaces and messages only?
2. If so, how efficient in terms of overhead is the command and control communication?
3. What is the trade-off between stealthiness of the command and control communication and the capacity of the hidden channel for conveying malicious information?

The rest of this paper is structured as follows.
In Section \ref{arch}, we give a short overview of the components and interfaces of 5G networks.
In Section \ref{design}, we discuss the attacker model and the design decisions.
In Section \ref{eval}, several possible attack scenarios using our framework are presented, and the design is evaluated against them.
In Section \ref{sec:countermeasures} we discuss possible countermeasures and why they are only partial solutions.
Section \ref{conclusion} summarizes our findings and presents future work.

\section{5G Core Network Architecture and Security}
\label{arch}

\begin{figure}
    \center
    \includegraphics[width=\linewidth]{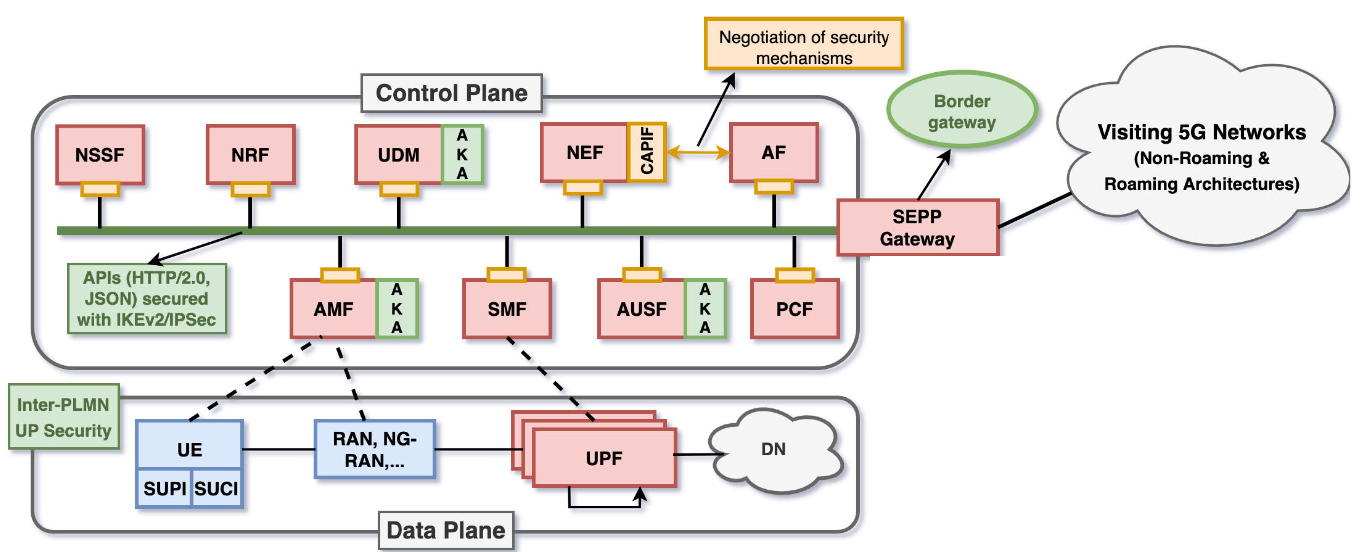}
    \caption{Overview of the 5G core network architecture \cite{TangSystematicAnalysis5G2022}.}
    \Description{Overview of the 5G core network architecture}
    \label{fig:5G_arch}
\end{figure}

In this section, a short introduction to the architecture of mobile networks in general and 5G in particular is provided.
Mobile networks can roughly be divided into three domains: \ac{ue}, \ac{ran} and core network.
While the \ac{ue} is typically understood as smartphones, many other devices can be considered \acp{ue} as well, e.g., vehicles, mobile Wi-Fi access points, smartwatches, and other smart home and smart city devices with cellular connectivity.
The \ac{ran} is the base station, consisting of radio hardware, signal processing equipment, and a so-called backhaul link (often fiber) to the core network.
The core contains the most diverse set of components that are organized as \acp{nf}.
The \ac{amf} is physically connected to the \ac{ran} and, via the \ac{ran}, logically to the \ac{ue}.
In 5G, the control and user plane are separated into different components, where the \ac{smf} controls one or more \acp{upf} using the \ac{pfcp}.
The \ac{upf} routes user data from the \ac{ran} to the \ac{dn}.
Based on the data transferred via the \ac{upf}, the \ac{pcf} provides the charging information for accounting.
Besides the \ac{ran} and \ac{upf}, the \ac{nef} connects to (external) \acp{af} and the \ac{sepp} connects to other mobile networks.
\acp{nf} communicate with each other via a \ac{sbi}.
The interface uses mutual \ac{tls} for confidentiality, authenticity, and integrity protection.
The \ac{nrf} provides service discovery and OAuth2.0 tokens for authorization between \acp{nf}.
The primary \ac{aka} for \ac{ue} security is initiated from the \ac{ue} via the \ac{amf}.
The \ac{amf} connects to the \ac{ausf}, which proxies the communication to the \ac{udm}.
The \ac{udm} stores long-term key material and location information for each \ac{ue}.
Interfaces that carry user data are referred to as u-plane and those that carry control data as c-plane.
An overview of the architecture is given in Figure \ref{fig:5G_arch}.

More details on the 5G core components can be found in 3GPP~TS~23.502, on 5G procedures in 3GPP~TS~23.502 and on security in 3GPP~TS~33.501.

\section{5G Puppeteer Design}
\label{design}
The overall objective of 5G Puppeteer is to provide stealthy command and control capabilities in a 5G core network.
To achieve this, the 5G Puppeteer system will send or piggyback onto standardized or genuine messages in the core network, which cannot be detected as malicious and filtered by intrusion detection systems.
Multiple command and control channels between single 5G core network components need to be established and chained together.
The chained channels represent the path which is used to convey information.
Paths can be divided into forward paths and backward paths.
Forward paths start at an entry point and are used to convey the information necessary to execute an attack (e.g., a \ac{supi}, \ac{pei} or \ac{ip} address) to the relevant component.
Backward paths are used to extract status information or attack result information via an exit point back to the attacker.
We describe the design in more detail in the following subsections.

\subsection{Attacker Model}
In our attacker model, we assume that the attacker has access to the 5G core components once to make modifications to the software of the components.
During the one-time access, the attacker installs the 5G Puppeteer system as backdoor with command and control capabilities on the 5G core network components.
The manipulated software is assumed to be able to read and modify incoming and outgoing 5G messages.

Even though this postulates a rather strong attacker, it is a realistic scenario as 5G networks are large, complex and partly outdated systems with various vulnerabilities and attack vectors.
Examples are supply chain attacks, malicious insiders, phishing attacks against the operators or maintenance staff, or exploitation of software bugs.

\subsection{Command and Control Channels}
The first step in the system design is to identify the available set of potential hidden channels between single 5G core components, i.e., the interfaces of the 5G network.
In general, two types of interfaces are relevant for 5G Puppeteer: External and internal interfaces.
External interfaces connect the 5G core components to external entities.
These interfaces are relevant for an attacker to send command and control messages from outside the core and to retrieve status messages or extracted information.
External interfaces considered in this paper are \ac{ue} to \ac{gnb} or \ac{amf}, and \ac{upf} to \ac{dn}.

Furthermore, there is an abundance of external interfaces available to design hidden command and control channels in addition to the ones considered in this work, examples being roaming (\ac{sepp} to \ac{sepp}), \ac{nef}, \ac{amf} to 4G \ac{mme}, and all other interfaces connecting the 5G core to legacy generation components as well as interfaces to the \ac{ims} and m-plane (i.e., management access for maintenance and manual configuration).

Internal interfaces connect two 5G core components within one \ac{plmn} with each other.
Typically, the 5G core relies heavily on the \ac{sba} which provides the \ac{sbi} for communication between two components.
The \ac{sbi} is using HTTP/2 and is protected with mTLS for confidentiality, integrity, and authentication. It uses the \ac{nrf} and OAuth2.0 for authorization.
In addition to the SBI, \ac{pfcp} is used to communicate between \ac{smf} and \ac{upf}, and \ac{ngap} is used for communication between \ac{gnb} and \ac{amf}.

Each internal and external interface has a set of messages that are standardized to be transferred over it.
The messages per interface are numerous and cannot be depicted here.
For an overview of \ac{sbi} messages, see 3GPP~TS~23.502 and for an overview of \ac{pfcp} messages see 3GPP~TS~29.244.
The abundance of possible messages and parameters in each message is a significant advantage for 5G Puppeteer as even if defense mechanisms are designed to detect command and control messages, the exact messages used for information embedding may vary from 5G Puppeteer deployment to deployment or over time.
Furthermore, any message can be part of the forward or backward path of the 5G Puppeteer system.

As covert channels, 5G Puppeteer overwrites used fields or uses optional fields.
From a steganographic point of view, these correspond to the hiding patterns E1n1 and E1.1n1 described in \cite{WendzelGenericTaxonomySteganography2025}.
However, many of the other patterns used in network steganography might be employed as well, ranging from embedding payload into redundancies of the network to modifying the inter-packet delays to transmit secret information.
For a comprehensive introduction to network steganography, we refer the reader to \cite{MazurczykNetworkSteganography2016}.

\subsection{Chaining Channels and Transient Messages}
\label{transient-msgs}
After a channel between two components is identified, information can be transmitted through it.
To execute sophisticated attacks in the network, the required command and control communication typically includes conveying messages over multiple components as the most sensitive components, e.g, the \ac{udm}, are often not exposed to external interfaces.
To achieve this, command and control channels can be chained.
Components that are part of the chain will retrieve command and control information from an incoming message and store it.
Then, they will send a message to the subsequent component in the command and control chain.
Furthermore, it is possible to tunnel (uncompromised) nodes in the 5G network to extend the command and control capabilities of 5G Puppeteer.
To achieve this, we take advantage of information being forwarded unmodified by a node as part of a genuine procedure.
For example, within the \ac{aka} sub-procedure, the \ac{suci} is generated by the \ac{ue} and sent unmodified to the \ac{udm}.
Such transient messages enable the forwarding of information even if not all components on the forward or backward paths are compromised.

\subsection{Information Embedding Modes and Message Routing}\label{Sec:Embedding_Routing}
To reduce the footprint and to circumvent detection and filtering, the covert channels of 5G Puppeteer only use connections between components for which genuine messages are defined.
Furthermore, 5G Puppeteer only uses standardized 5G messages to transmit hidden data.
Moreover, the messages used are part of a complete 5G procedure that involves the attacker, e.g., the registration procedure or bearer establishment procedure.
These three rules ensure that the hidden command and control channels are hard to distinguish from genuine 5G procedures.
The mode in which the 5G Puppeteer system follows these three principles is referred to as \ac{im3c}.

An optional goal is to not require any additional messages for command and control.
5G Puppeteer can archive this by piggybacking the desired command and control information onto genuine messages and procedures, thus further reducing the risk of detection and filtering.
This mode is referred to as \ac{pb3c}.
In the \ac{pb3c} mode, special care has to be taken in case multiple paths exist in the network for building a command and control channel.
As upcoming procedures cannot be foreseen but only estimated, we do not know the fastest path in advance.

Thus, we propose three simple routing options that do not assume a-priori knowledge of the network topology (i.e., possible paths as provided by protocols such as the Spanning Tree Protocol (STP) in Ethernet networks):
\begin{enumerate}
    \item Path Flooding (PF):
          The fasted option, which also provides some redundancy, is to send the messages on all available paths.
          The feasibility of this approach depends on the number of command and control messages expected.
          If 5G Puppeteer is used to trigger attack execution,           this option will be feasible.
          On the other hand, if 5G Puppeteer is used to regularly exfiltrate status updates or u-plane data, i.e.,           network traffic of a target,           this option might cause problems in both \ac{im3c} (due to a significant number of additional messages) and \ac{pb3c} (due to high latencies) modes.
    \item Round Robin (RR):
          Round Robin uses all suitable paths in turn. This reduces the overall number of messages but provides no redundancy and the chosen path might not be the fastest.
    \item Estimate-Enhanced Round Robin (EE-RR):
          This variant uses all suitable paths based on weights.
          The weights represent the estimated speed of a path.
          Estimation of optimal weights depends on the expected or monitored network load, the frequency of specific procedures, and the set of compromised nodes that can be used for command and control.
          Additional adjusting of the weights is possible, e.g.,           to reduce load on a specific component to facilitate stealthiness.
\end{enumerate}

These considerations are specific to the \ac{pb3c} mode, as the messages in the \ac{im3c} mode are created and sent whenever necessary and, therefore, no delay other than the regular network latency is expected.
However, one path must be selected among all available paths for an attack in the \ac{im3c} mode, too.
This path may be chosen based on stealthiness or performance considerations, e.g., minimizing the load on a particular component or the number of procedures required.
In general, the \ac{pb3c} mode is stealthier and the \ac{im3c} mode is faster and more deterministic in terms of the timing of the attack execution.

\subsection{Information Embedding: 5GPP}
The embedding of command and control messages can be achieved in various ways.
In this paper, we exclude options for establishing hidden channels and any protocols that are not 5G-specific, i.e., \ac{http}, \ac{tls}, \ac{ip} and other lower layer protocols as well as the \ac{udp} stack that supports \ac{pfcp}.
Instead, we focus on the registration procedure as it is used rather frequently in any 5G deployment.
Specific attacks or an increased performance might require the consideration of further procedures or messages.
Other (5G) protocols, e.g., \ac{rrc}, \ac{pdcp} or \ac{gtp-c} might be suitable to embed command and control messages as well.
The following subsections contain a structured analysis of the embedding capacity of network messages that are sent and received on different interfaces as well as a light-weight protocol to be used for command and control communication.

\subsubsection{External interfaces}
The \ac{upf} and the \ac{ue} components are used as entry and exit interfaces, meaning that malicious information can be entered or exfiltrated through them.
The \ac{upf} includes an u-plane interface to the \ac{dn}, in particular to the Internet.
Therefore, a \ac{upf} can freely interchange information with external entities under the control of an attacker, e.g., any website or a \ac{ftp} server.
\acp{ue} can be a simple modem with an attached \ac{sim} to connect to the 5G network.
If a modification of the c-plane is required to embed information in lower layer protocols, e.g., in the \ac{rrc} protocol, open-source software like srsRAN \cite{srsRAN} or OpenAirInterface \cite{OpenAirInterface} can be used to transmit and receive 5G messages with arbitrary modifications to the protocol stack.
Therefore, we consider any information available at one of the interfaces to be available to an attacker for extraction.

\subsubsection{Service-based Interfaces}
The \acp{sbi} contain plenty of options to embed bits into the network messages.
Depending on the mode used, the possibilities of the attacker vary:
In the \ac{im3c} mode, an attacker might alter any parameter available allowing for significant command and control capabilities.
Even if only optional or insignificant parameters are changed in the \ac{pb3c} mode, the potential for command and control is significant on the \ac{sbi} as well.
An example is the \texttt{Nsmf\_PDUSession\_Create} message that is part of the registration procedure and used to inform the \ac{smf} of relevant parameters to provide session management and configure the \ac{upf} properly.
It contains 21 optional parameters, some of which are whole containers with many further parameters.
The response to this request contains up to 8 parameters, which also include further containers.
Table~\ref{tab:attack_overview} provides an overview of the capacity of the \ac{sbi} messages that are part of the registration procedure for the embedding of malicious information.
\begin{table*}[ht!]
    \centering
    \caption{Overview of the capacity of the SBI messages that are part of the registration procedure for the embedding of malicious information. \smallskip}
    \begin{tabularx}{0.98\textwidth} { |l| *{4}{Y|} }
        \hline
        \multirow{2}{*}{\bf{Message}}          & \multicolumn{2}{c|}{\;\bf{Request Parameters\;}} & \multicolumn{2}{c|}{\bf{\;Response Parameters\;}}                       \\ \cline{2-5}
                                               & Required                                         & Optional                                          & Required & Optional \\ \hline
        \texttt{Nudm\_UECM\_Registration}      & 7                                                & 2                                                 & 1        & -        \\ \hline
        \texttt{Nudm\_SDM\_Get}                & 3                                                & 1                                                 & 1        & -        \\ \hline
        \texttt{Nudm\_SDM\_Subscribe}          & 2                                                & 1                                                 & -        & -        \\ \hline
        \texttt{Npcf\_AMPolicyControl\_Create} & 11                                               & -                                                 & 2        & 5        \\ \hline
        \texttt{Nsmf\_PDUSession\_Create}      & 12                                               & 6                                                 & 10       & 6        \\ \hline
        \texttt{Nudm\_SDM\_Info}               & 3                                                & 1                                                 & 1        & -        \\ \hline
    \end{tabularx}
    \label{tab:attack_overview}
\end{table*}

Note that the number of parameters is only an indication of the available capacity for command and control:
A parameter can be a single value as well as a container including further parameters or even further containers.
Additionally, single values might be anything in between a 1-bit information or an \ac{ip}v6 address consisting of 128 bit.
Thus, there are various options to embed information into genuine \ac{sbi} messages catering to different performance, latency, and stealthiness requirements.

\subsubsection{Uu, N2, and N1 Interfaces}
The Uu radio interface physically connects the \ac{ue} to the \ac{gnb} and the \ac{rrc} protocol is the most significant protocol for communication.
The initial registration request of a \ac{ue} includes several hundreds of optional parameters that might be abused for command and control communication.
Other messages included in the \ac{as} \ac{smc} procedure might be used for bidirectional communication as well.
The N2 interface connects the \ac{gnb} to the \ac{amf}.
In addition to the required messages for configuration between \ac{ran} and 5G core, this interface is used to relay messages between the \ac{ue} and the \ac{amf}.
For example, the \ac{ngap} protocol might be used to embed information
to execute attacks on the \ac{gnb}.
The N1 interface logically connects the \ac{ue} to the \ac{amf} and is using the \ac{nas} protocol.
The \ac{nas} protocol contains various messages that could be used to embed secret information, for example the forwarded registration request.
Other messages, like the \ac{nas} \ac{smc}, the identity request, and response messages as part of the registration procedure, can be used as well.

\subsubsection{N4 Interface}
The N4 interface between the \ac{smf} and \ac{upf} uses the \ac{pfcp} for communication.
There are 14 procedures with a total of 24 messages defined for this interface, including periodic heartbeat messages.
Therefore, there are several options for the establishment of command and control communication on the N4 interface.
As in this work the emphasis is on the registration procedure, the relevant messages are session management-related messages, namely the \ac{pfcp} Session Establishment Request and the \ac{pfcp} Session Establishment Response.
The \ac{pfcp} Session Establishment Request message is initiated by the \ac{smf} during the registration procedure.
It includes 26 mandatory and optional parameters.
The \ac{upf} responds with the \ac{pfcp} Session Establishment Response message, which includes 14 mandatory and optional parameters.
If the \ac{upf} needs to initiate the communication, the \ac{pfcp} Session Report Request message can be used to embed secret information.
This message includes 12 parameters including containers usable for hidden information transfer and is answered by the \ac{pfcp} Session Report Response message, which includes seven parameters some of which are containers with further parameters.

\subsubsection{5G Puppeteer Protocol}
In general, the protocol to transfer information between the compromised nodes can be optimized for various scenarios, e.g. the supported type of attacks.
However, we propose the lightweight protocol \ac{5gpp} that provides basic network protocol properties to enable command and control communication.
For the design of \ac{5gpp}, the following requirements are considered:
\begin{itemize}
    \item R1 Security: To increase the difficulty to detect \ac{5gpp} and to increase complexity in analysis of the command and control messages, parameters are encrypted. However, the integrity, availability, and authenticity of the messages are not addressed.
    \item R2 Routing: Depending on the scenario, different routing options are available to allow for dynamic optimization of the command and control messaging, e.g., for the trade-off between attack execution speed and stealthiness.
    \item R3 Concurrency: It is possible to execute multiple attacks in parallel and to receive the results of them.
    \item R4 Attack types: A number of different attacks are supported.
    \item R5 Dynamic paths: The exit node and, if possible, the execution node are configurable in the protocol to allow for a more fine-grained attack scenarios.
          The entry node is implicitly chosen to be the first node receiving the command and control messages.
    \item R6 Message splitting: It is possible to split large payload into multiple messages to allow command and control messages also in cases where only little space for information embedding is available in genuine messages.
    \item R7 Forwarding limitations: Each message is only forwarded a limited number of times to prevent infinite loops that would increase the network load and decrease the stealthiness of the command and control communication.
\end{itemize}

The \ac{5gpp} uses a 20-bit header and a dynamic payload size, depending on the executed attack.
Table \ref{tab:5gpp_header} shows the \ac{5gpp} header structure.

\begin{table}[ht!]
    \centering
    \caption{Structure of the 5GPP header. \smallskip}
    \begin{tabular}{ |c|l|l|l| }
        \hline
        \bf{\phantom{a}Bit \phantom{a}} & \bf{Type \phantom{aaa}} & \bf{Interpretation \phantom{aaaaaaaaaaaaaaa}} & \bf{Security \phantom{aaa}} \\ \hline
        1-4                             & Key ID                  & Sym. key for encryption                       & Open                        \\ \hline
        5-6                             & Routing option          & PF: 1, RR: 2, RR-W: 3, etc.                   & Open                        \\ \hline
        7-9                             & TTL                     & Limit of hops                                 & Open                        \\ \hline
        10                              & Splitting indication    & Not split: 1, Split: 2                        & Encrypted                   \\ \hline
        11-13                           & Execution point         & UDM: 1, AMF: 2, etc.                          & Encrypted                   \\ \hline
        14-16                           & Attack ID               & Up to 8 synchronized attacks                  & Encrypted                   \\ \hline
        17-18                           & Attack type             & UDM key ext.: 1, PWS: 2, etc.                 & Encrypted                   \\ \hline
        19-20                           & Exit point              & UE: 1, UPF: 2, SEPP: 3, NEF: 4                & Encrypted                   \\ \hline
    \end{tabular}
    \label{tab:5gpp_header}
\end{table}

The security of \ac{5gpp} is provided by a symmetric key which is deployed by the attacker during the compromise of a node.
Up to 16 different keys can be deployed so that only the relevant node or group of nodes is able to decrypt the attack parameters.
It is important to prevent the disclosure of execution and exit nodes as they can leak information about which other nodes are compromised.
For routing, the options discussed in Section~\ref{Sec:Embedding_Routing} are available.
The Time-To-Live (TTL) is a value $n \in \mathbb{N}$ which is decremented in each node prior to the forwarding of the message and, after a maximum of $n$ nodes, the message is dropped to prevent infinite loops.

We assume that an attacker configures the compromised nodes to retrieve the \ac{5gpp} header from well-known messages and parameters in the message.
In general, if the compromised nodes do not have information on the location of embedded information, a magic number (e.g.\ 4~bit) in the beginning of the header and a checksum to validate the message content (e.g.\ 6~bit) at the end of the header are sufficient for identification and validation of the malicious information.

\section{Analysis and Evaluation}
\label{eval}
For analysis and evaluation, only a subset of the discussed 5G Puppeteer capabilities are simulated.
We implement the \ac{pb3c} mode with the Path Flooding routing.
We use the registration procedure, including identity request and response, \ac{aka} and \ac{smc} as given in Figure~\ref{fig:registration}.
This setup excludes N32/\ac{sepp} as well as \ac{nef} entry and exit nodes and limits the overall set of components reachable in the network to: \ac{ue}, \ac{gnb}, \ac{amf}, \ac{ausf}, \ac{udm}, \ac{smf}, and \ac{upf}.
Nevertheless, several attacks are executable with these components.
The simulation is implemented in Python with networkx and matplotlib for visualization.

\subsection{Environment Representation}

The 5G core environment is represented by the tuple $E$ defined as
\begin{equation} \label{EQ:Def_E}
    E = (Procedures, Compromised Nodes).
\end{equation}

The set $Compromised Nodes$ determines which 5G components will store and forward malicious information.

$Procedures$ is the ordered set of all procedures of the 5G network.
Each procedure is a tuple containing all messages of the procedure in the order they will be sent.
A single message $m$ is given by
\begin{equation} \label{EQ:Def_m}
    m = (Source, Target, Available Space), \end{equation}

where $Source$ is the 5G core component that is sending the message, $Target$ the receiving component, and $Available Space$ indicates the number of bits that can be used for embedding payload into the message.

The order of messages in the procedures is crucial to the effectiveness and efficiency of 5G Puppeteer, as information can only be forwarded within a particular message, if it arrived at the sending node before this message is sent.
Thus, the choice of the procedure for the attack has a huge impact on the available paths for the attack and the temporal behavior.

\begin{figure}
    \center
    \includegraphics[width=\linewidth]{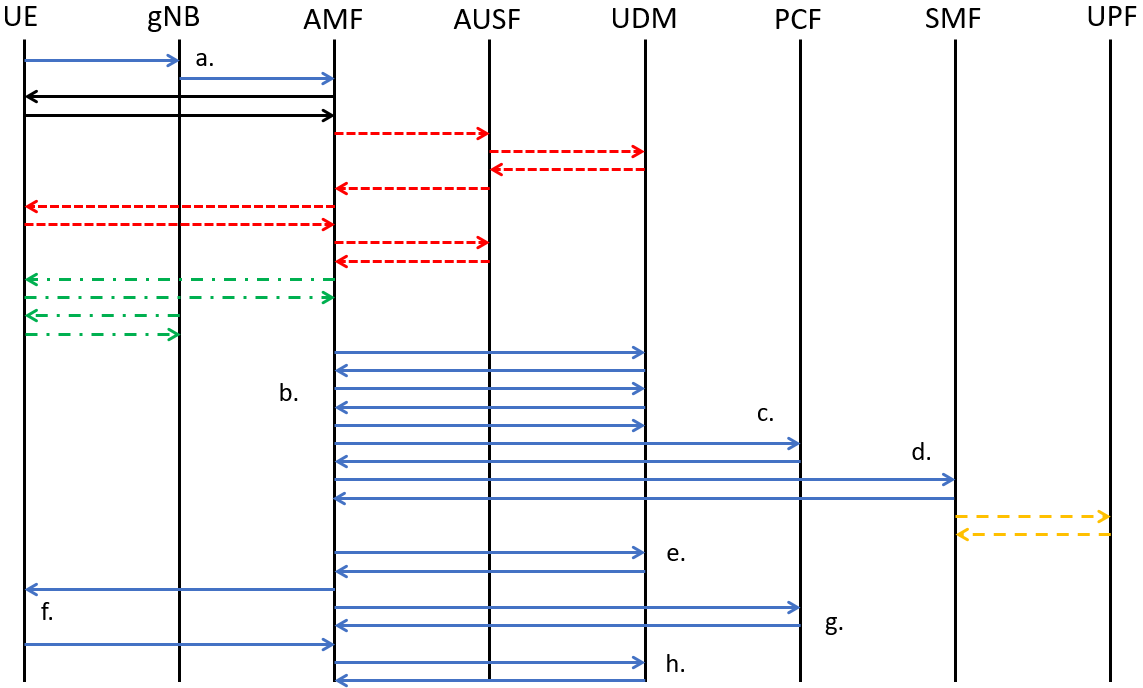}
    \caption{Signaling diagram for the implemented registration procedure. Blue solid: Registration procedure, black solid: identity sub-procedure, red dashed: AKA sub-procedure, green dash-dotted: SMC sub-procedure, yellow dashed: N4 u-plane configuration. Based on \cite{ZeidlerPerformanceEvaluationTransport2024}}
    \Description{Signaling diagram for the implemented registration procedure. Blue solid: Registration procedure, black solid: identity sub-procedure, red dashed: AKA sub-procedure, green dash-dotted: SMC sub-procedure, yellow dashed: N4 u-plane configuration}
    \label{fig:registration}
\end{figure}

In this work, we use the registration procedure and implement it as shown in Figure \ref{fig:registration}.
Note that the implementation of this procedure might differ (slightly) between mobile networks.
Our implementation works as follows:

The registration is initiated by the \ac{ue} registration message (a.), which is received by the \ac{gnb} and forwarded to the \ac{amf}.
The \ac{amf} requests the identity of the \ac{ue} (black solid).
Afterwards, the \ac{amf} initiates the \ac{aka} sub-procedure (red dashed) to establish a security context, followed by the \ac{smc} sub-procedure (green dash-dotted) after success to enable message protection.
The \ac{amf} then updates the \ac{udm} (b.), \ac{pcf} (c.) and \ac{smf} (d.) with status information of the registration to enable data routing and charging.
Data connectivity is activated after the \ac{smf} configures the \ac{upf} with the necessary information received from the \ac{amf} (yellow dashed).
Finally, the \ac{amf} confirms the registration with the \ac{udm} (e.), the \ac{ue} (f.), the \ac{pcf} (g.), and, after \ac{ue} re-confirmation, the \ac{udm} (h.).

\subsection{Attack Representation}\label{Sec_Attacks}
An attack $A$ is represented as tuple
\begin{equation}
    A = (Entry, Execution, Exit, Forward Bits, Backward Bits).
\end{equation}
Here, $Entry$ is the node at which the malicious information enters the 5G Puppeteer network.
As described, possible entries are the \ac{upf} and the \ac{ue}.
The $Execution$ parameter defines the component that is able to execute the intended attack.
If the malicious information reaches the executing node, the attack is completed.
The execution node can also abuse genuine procedures of other components to execute the attack, e.g. triggering the localization procedure to obtain the location of a subscriber.
After the attack is completed, the malicious information, e.g., the extracted information or the attack status, traverses the 5G Puppeteer network back to the $Exit$ node.
$Forward Bits$ depicts the number of bits required at the execution node to perform the attack and $Backward Bits$ the number of bits that need to be conveyed back to the exit node.

If core network components are compromised, an abundance of attacks is possible.
In the subsequent paragraphs, we describe three exemplary attacks A1, A2, and A3 to showcase the overall concept and design choices to be made.
Table \ref{tab:attack__param_overview} lists the parameters for the attacks A1 to A3.
An overview of further attacks in mobile networks is provided in \cite{AhmadOverview5GSecurity2018}.

\begin{table}[ht!]
    \centering
    \caption{Overview of the exemplary attack parameters used for simulation}
    \label{tab:attack__param_overview}
    \begin{tabular}{ |l|c|c|c|c|c| }
        \hline
        \multirow{2}{*}{\bf{Attack}}   & \multicolumn{3}{c|}{\bf{Attack path}} & \multicolumn{2}{c|}{\bf{Attack execution (Bits)}}                                                                 \\ \cline{2-6}
                                       & Entry \phantom{a}                     & Exec. \phantom{a}                                 & Exit \phantom{a} & Forward \phantom{a} & Backward \phantom{a} \\ \hline
        A1: Key Extraction \phantom{a} & UE                                    & UDM                                               & UE               & 128                 & 192                  \\ \hline
        A2: UE Localization            & UPF                                   & AMF                                               & UE               & 96                  & 112                  \\ \hline
        A3: PWS Abuse                  & UE                                    & gNB                                               & -                & 32                  & -                    \\ \hline
    \end{tabular}
\end{table}

\subsubsection{A1: Extraction of Long-Term Key Material}
Mobile network security relies heavily on symmetric cryptography.
The symmetric long-term key $k$ is stored on the \ac{uicc} on the end-user side and in the \ac{udm} in the 5G core network.
Session keys (i.e., $k_{gNB}, k_{SEAF}, k_{AUSF}, k_{AMF}$, for details see TS~33.501, Annex~A) are derived from the long-term key and additional information, which can easily be obtained by passive radio monitoring.
Therefore, the security of 5G relies on the security of $k$.
If $k$ is compromised, the user data, e.g., internet traffic, are vulnerable to wiretapping.

In this attack scenario, an attacker sends a target identifier, e.g., the \ac{supi}, the \ac{suci}, or the \ac{pei}, via an entry point to the 5G core network.
The network conveys this information to the compromised \ac{udm}.
The \ac{udm} extracts $k$ for the corresponding \ac{supi} and sends it to an exit point of the network.
The attacker receives $k$ at the exit point.

This basic example of the capabilities of 5G Puppeteer with only a single compromised function showcases the extraction of information deep within the 5G core network based on parameters, i.e., target information, from the core boundary.

\subsubsection{A2: Obtaining Location Information from Users}
The location of a user (or the \ac{ue}) is highly sensitive information.
Compromising the location of a user might be used for user tracking and behavior analysis.
In 5G, the \ac{udm} holds limited user location information, however, the \ac{lmf} is able to determine the user position with an accuracy of down to several meters in practice and, in theory, few centimeters \cite{PeralRosadoSurveyCellularMobile2018}.
Therefore, the \ac{lmf} is a valuable target for attackers.
In this attack scenario, the attacker again sends target information via an entry point into the network.
The network conveys the information to the compromised \ac{amf}, which requests the location information from the \ac{lmf} and exfiltrates the location information via an exit point.
Interestingly, the \ac{amf} does not have access to the location information of a user itself.
Instead, the \ac{amf} needs to abuse the genuine localization procedure on behalf of the attacker for the localization of the target.
This is an example for a combination of malicious command and control communication and genuine 5G core network procedures to achieve the objectives of the attack.

\subsubsection{A3: Abusing the Public Warning System}
In the \ac{pws}, a.k.a.\ cell broadcasting, a \ac{gnb} sends messages to alert all \acp{ue} in a particular cell.
For this, it uses \ac{sib} messages that are frequently broadcast in each cell to facilitate cell selection and re-selection.
General security problems with the authenticity in the \ac{pws} are discussed in \cite{BitsikasYouhavebeen2022}.
The \ac{pws} adds additional \ac{sib} messages, i.e., SIB6, SIB7, SIB8, that are received and processed by \acp{ue} in the cell and that will trigger the alert.
False alerts in crowded areas may cause panic and are therefore a physical security risk.
In this scenario, an attacker sends no target information, but instead the type of alert to be triggered and, optionally, an alert message shown on the display of alerted \acp{ue}.
This showcases that there are attacks without target information and, more interestingly, that 5G Puppeteer can be easily extended to include command and control capabilities in the \ac{ran}, i.e., a \ac{gnb}, as well.
\enlargethispage{1\baselineskip}

\subsection{Attack Success and Path Selection}
Based on the chosen environment $E$ and attacks $A$, we can determine the requirements for an attack to be successful.
The following method finds paths available for a 5G Puppeteer orchestration:

First, identify the set of compromised core components.
Second, identify all procedures that include messages between at least two of the compromised components.
Third, identify all messages within these procedures.
Fourth, verify for each of these messages that at least 1 bit of command and control information can be embedded in the message.
This step is optional in the \ac{im3c} mode as any information in the message can be modified, because the message is induced by the attacker and not needed for genuine operation of the network.
Fifth, build a network connectivity graph from this.
In this graph, the network components are depicted as nodes and messages are directed edges.
This is the 5G Puppeteer network.

Between any two 5G core network components there can be both a forward and a backward path, only one of the two, or no path at all.
This needs to be considered to determine the set of possible attacks in the network.
Whether an attack can be executed in the 5G Puppeteer network can be examined by basic graph traversal algorithms like depth-first search (DFS) or breadth-first search (BFS).

\begin{figure}[ht]
    \center
    \includegraphics[width=\linewidth]{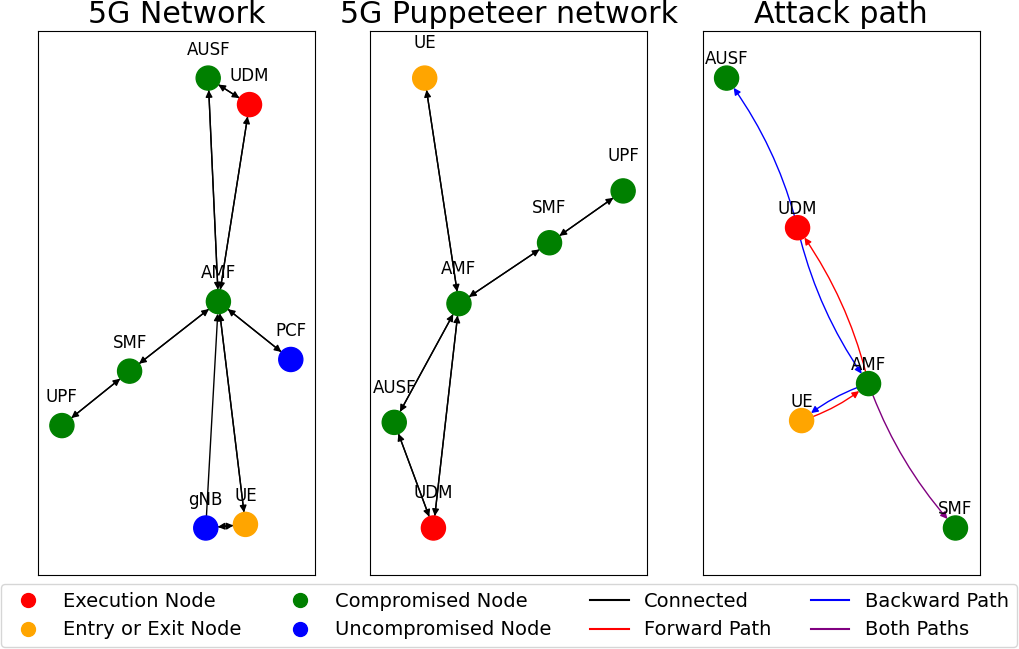}
    \caption{Simulation result of the attack success and path selection of 5G Puppeteer visualized in our simulation tool with the networkx library. Left: Overall 5G Network that is available with messages send. Center: Part of the 5G network that is available to 5G Puppeteer. Right: Actual attack path selected.} \label{fig:attack_success}
    \Description{Simulation result of the attack success and path selection of 5G Puppeteer visualized in our simulation tool with the networkx library. Left: Overall 5G Network that is available with messages send. Center: Part of the 5G network that is available to 5G Puppeteer. Right: Actual attack path selected.}
\end{figure}

Here, we consider the registration procedure and visualize the approach in Figure~\ref{fig:attack_success}:
The left part of the figure visualizes the overall part of the 5G core network that is reachable by the messages included in $E$.
Therefore, this network represents possible execution, entry, and exit nodes.
For the registration procedure, the components included are:
\ac{ue}, \ac{gnb}, \ac{amf}, \ac{smf}, \ac{pcf}, \ac{upf}, \ac{ausf} and \ac{udm}.
With this, the most significant components are present in the 5G Puppeteer network built upon the registration procedure.
The middle pane of the Figure~\ref{fig:attack_success} visualizes the reduced 5G Puppeteer network that remains, if the command and control message path begins at the entry point \ac{ue}, and only compromised components can receive, store and forward malicious information.
This limitation could be worked around using \textit{transient messaging}, as discussed in Section \ref{transient-msgs}.
On the right part of the figure, the actual attack path starting at  the entry point (\ac{ue}, orange node) and finishing at the exit point (\ac{ue}, orange node) can be seen.
The directions of the edges indicate the information flow and their colors whether it is an information flow in the forward (red), backward (blue) or both (purple) paths.
For the routing of the messages, we used the Path Flooding algorithm proposed in Section~\ref{Sec:Embedding_Routing}.
The attack showcased is A1, the \ac{udm} key extraction attack with \ac{supi} target (64 bit), \ac{supi} exit point (64 bit) and \ac{5gpp} header (20 bit).
Each message in the registration procedure was assumed to have a capacity of 64 bit for command and control information to be embedded.
In this example, the forward and the backward paths are similar (\ac{ue}~$\leftrightarrow$~\ac{amf}~$\leftrightarrow$~\ac{udm}).
Due to the path flooding, the information on the forward path is also conveyed to the \ac{ausf} and on the backward path to the \ac{smf}.
If the key IDs in \ac{5gpp} are chosen well, neither of these components would be able to decrypt the attack information.
This scenario also showcases the need for the time-to-live parameter in \ac{5gpp} to prevent infinite looping of the command and control messages.
The \ac{udm} key extraction attack can be performed within four consecutive registration procedures in the \ac{pb3c} mode with path flooding as routing algorithm.
For given $E$ and $A$, our developed 5G Puppeteer simulation tool determines whether the attack can be successful in general (center part of the figure) and, if so, the optimal forward and backward path as well as the number of specific procedures required for attack completion.

\subsection{5G Puppeteer Performance}
\begin{figure}
    \center
    \includegraphics[width=\linewidth]{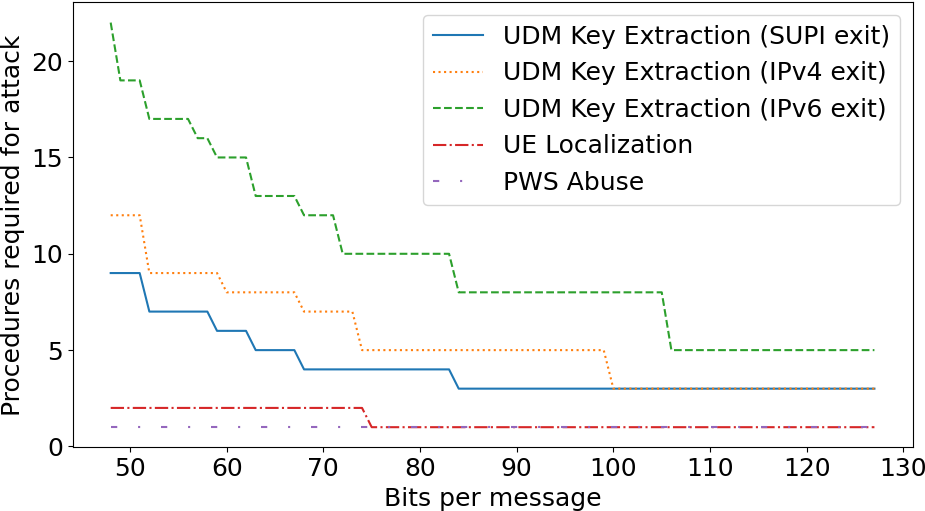}
    \caption{Performance of exemplary attacks at different payload sizes.} \label{fig:5gp_performance}
    \Description{Performance of exemplary attacks at different payload sizes.}
\end{figure}

The developed 5G Puppeteer simulation tool is also able to alter the used space within messages to analyze the impact of more aggressive information embedding, i.e., more information per message, which is a trade-off with respect to stealthiness.
Furthermore, the tool can be used to evaluate, for given $E$ and $A$, the requirements for embedding information to fulfill the attacker's objectives.
Figure \ref{fig:5gp_performance} presents an overview for the attacks
A1, A1 with IPv4 exit point, A1 with IPv6 exit point, A2, and A3 as introduced in Section~\ref{Sec_Attacks}.
We assume that \ac{amf}, \ac{udm}, \ac{smf}, \ac{udm}, \ac{gnb}, and \ac{ausf} were compromised and \ac{ue}s were used as entry and exit point.
For A1 with IPv4 and IPv6 exit points, the \ac{upf} was used as exit node.

For a low number of bits embedded in messages, some attacks require up to 22 registration procedures to transport the information as required by the attack.
If only the minimal number of bits (21 bit) is available, some attacks require up to 612 registration procedures for attack completion.
On the other hand, for embeddings of 48 or more bits per message most attacks can be completed in less than 15 procedures (see Figure~\ref{fig:5gp_performance}).
As registration procedures occur rather regularly (for initial registration, re-registration after a specified period of time (Timer T3512), and after the security context or the \ac{guti} cannot be mapped) in 5G networks, we estimate that any attack can be completed in seconds.

\subsection{Overhead Analysis of the 5G Puppeteer Protocol}

The \ac{5gpp} provides several basic properties to facilitate routing, confidentiality, dynamic attack configuration, etc.
However, the 20 bit \ac{5gpp} header introduces an overhead into the 5G Puppeteer command and control communication.
Table~\ref{tab:5gpp_example_overhead} provides some example headers for the attacks A1-3.
Most parameters are not attack specific but scenario specific and can be adjusted as required, e.g., the exit point or time-to-live.
The overhead of the header is discussed for the three examples in the following paragraphs.

\begin{table}[ht!]
    \centering
    \caption{Example 5GPP headers for different attack types.}
    \begin{tabular}{ |c|l|c|c|c| }
        \hline
        \bf{Bit} & \bf{Type}            & \bf{A1 (SUPI exit)} & \bf{A1 (IP exit)} & \bf{A3 (PWS)} \\ \hline
        1-4      & Key ID               & Any (1-16)          & Any (1-16)        & Any (1-16)    \\ \hline
        5-6      & Routing option       & Any (1-4)           & Any (1-4)         & Any (1-4)     \\ \hline
        7-9      & Time-to-live         & Any (1-8)           & Any (1-8)         & Any (1-8)     \\ \hline
        10       & Splitting indication & Any (1-2)           & Any (1-2)         & Any (1-2)     \\ \hline
        11-13    & Execution point      & 1                   & 1                 & 2             \\ \hline
        14-16    & Attack ID            & Any (1-8)           & Any (1-8)         & Any (1-8)     \\ \hline
        17-18    & Attack type          & 1                   & 1                 & 2             \\ \hline
        19-20    & Exit point           & 1                   & 2                 & 1             \\ \hline
    \end{tabular}
    \label{tab:5gpp_example_overhead}
\end{table}

\textbf{A1 (SUPI exit point)}:
The payload consists of the target SUPI and the SUPI of the UE which acts as the exit point.
Each \ac{supi} has a length of 64 bit, thus the 20 bit header would lead to an overhead of $20/128 = 15.6~\%$ in the forward path and, as the extracted key is 128 bit as well, $20/(128 + 64) = 10.4~\%$ in the backward path.
The maximum packet size on the backward path is $212~\text{bit}$.

\textbf{A1 (IP exit point)}:
The payload consists of the target SUPI and the IP address of the exit point.
This is either a 32 bit IPv4 address or a 128 bit IPv6 address resulting in a forward path overhead of $20.8~\%$ or $10.4~\%$, respectively.
The overall packet size in the forward path is $116~\text{bit}$ (IPv4) or $212~\text{bit}$ (IPv6).
The backward path has an overhead of $20/(128 + 32 + 20) = 11.1~\%$ (IPv4) or $20/276 = 7.2~\%$ (IPv6), respectively.

\textbf{A3 (PWS)}:
The payload is the target cell ID, in which the PWS shall be activated.
Since the exit point is not used, only the cell identifier ($22$--$32$ bit) is needed, resulting in an overhead of $62$--$90~\%$ and a total packet size of $42$--$52~\text{bit}$ depending on the cell identifier.
The PWS might also be abused to send specific text messages, this would require further space to embed the type of warning message and the text message itself.
One could easily include this attack in the \ac{5gpp} by defining a further attack type in the protocol.

In conclusion, the 20 bit overhead is not significant in general but in certain scenarios with small payload, e.g., the PWS abuse attack, it can reach up to $90~\%$.
As the overall message size in these cases is rather small, e.g., only 42 bits in the case of the PWS abuse attack, this is not a critical issue.
We also conclude that $276~\text{bit}$ of space to embed information in a message are sufficient for even the largest payloads for the evaluated attacks and, if a lower number of bits is available, the payload can be split among multiple messages.
When splitting the payload, only 21 bits of space are required in each message ($20$ for the header, $1$ for the payload), resulting in a maximal overhead of $20/1 = 2000~\%$.
However, this is unlikely, as typically more space is available to embed information.

\subsection{Transient Messaging Attacks: An AKA Case Study}
The \ac{5gpp} provides several basic networking protocol properties.
However, for extremely optimized attacks, other protocols or messages can be defined.
An example for an almost undetectable execution of the \ac{udm} key extraction attack in \ac{aka} might work as follows:
The \ac{ue} can bridge the \ac{amf} and \ac{ausf} because both will forward the unmodified \ac{suci} provided by the \ac{ue} until it reaches the \ac{udm}.
Therefore, only the \ac{udm} needs to be compromised.
During compromise, the attacker stores a symmetric key $k_{attack}$ in the \ac{udm}.
The \ac{mac} tag included in the \ac{suci} has 64 bit.
The attacker encrypts, e.g., with the \ac{aes}, the target \ac{supi} (64 bit) with the key $k_{attack}$ and embeds the cipher text into the \ac{mac} tag parameter in the \ac{suci}.
If the \ac{udm} receives a registration request from the attacker's \ac{supi}, it will perform the registration as requested but instead of verifying the \ac{mac} tag for integrity protection of the message, $k_{attack}$ is used to decrypt the \ac{mac} tag to extract the target \ac{supi}.

As a backward channel with transient properties, the authentication request can be used, where the RAND (128 bit) and AUTN (128 bit) parameters generated in the \ac{udm} are forwarded unmodified up to the \ac{ue}.
The two parameters make up 256 bit and can therefore convey two extracted keys (128 bit each) at once.
To prevent detection, the extracted keys are encrypted with \ac{aes} using $k_{attack}$.
The attacker can decrypt the extracted keys received at the \ac{ue} with $k_{attack}$ to retrieve the symmetric long-term key $k$ of the target \ac{supi}.
\enlargethispage{1\baselineskip}

This method allows for an almost undetectable key extraction as the registration procedure is completed in a regular fashion and the information is embedded into high entropy parameters while the information has also a high entropy due to the encryption.

\section{Countermeasures}
\label{sec:countermeasures}

As shown in the previous section, this hidden communication is very hard to detect.
Especially when transient messages are used, the lack of verifiability of actions by critical components such as the \ac{udm} leads to a large attack surface.
Based on these assumptions we propose two possible detection approaches: Specific behavior analysis and shadow functions.
Both approaches will be discussed in this section.

\subsection{Specific Behavior Analysis}

While many existing intrusion detection systems perform anomaly detection on individual \acp{nf}, in our opinion, these are too broad to detect single attacks of this type.
Especially transient paths only require a single or a few messages and will therefore stay below the noise floor.
These attacks most likely require a fine-tuned detection algorithm that is able to operate on a per instruction level.
For the previously introduced \ac{udm} example, this could be a trained system that audits the usage of the \ac{aesni} instruction set.
This assumes that the \ac{udm} is a piece of software that is run on operator controlled infrastructure.
While the software itself remains a black box, the operator could train a system using known safe requests to associate each invocation of \ac{aesni} with a given source function.
Based on this training information, any deviation from this in production could trigger an alert and prevent the execution.

While this approach should be able to detect these kinds of attacks, an adversary could harden its software if the exact operation is known.
In this example, the malicious code could sometimes execute the malicious internal function while dropping its result and therefore adding it to the allow-list without triggering a failure in testing.
Additionally, each new software version would require extensive re-training, possibly delaying security fixes and introducing additional costs.

\subsection{Shadow Functions}

The other approach is similar to the shadow stack employed by modern processors.
Instead of buying equipment from only a single, possibly untrusted, vendor, an additional instance of the function is procured from a different vendor.
If both vendors are carefully selected (e.g., different spheres of influence, different supply chains, etc.) the chance of both being manipulated in the same way is significantly reduced.
The single instance can therefore be used as a means of sanity checking the production functions.
This check involves the duplication of a certain percentage of all requests and forwarding the duplicated message to the independent shadow instance.
If this instance has the same state as the production instances, for most requests the output should be identical or similar.
If the response of the shadow instances deviates unexpectedly from the production function, this could trigger an alert and block the response.

Similar to the previous approach, this also comes with certain limitations.
Depending on which function should be shadowed, synchronizing the state can be simple to complex.
Whenever randomness is involved, the shadow function would also need a way to re-use the values selected by the primary instance, likely requiring support from both vendors.
Additionally, due to different interpretations of the standard, different vendors likely respond differently to the same request, again requiring a training phase with known good requests and responses.
Other than the previous approach, these differences are unlikely to change with each subsequent software version, reducing the training effort required.

While both approaches are not perfect, they should in theory allow detection of this kind of attacks.
How practical they are depends on the exact system a network operator has.

\section{Conclusion and Future Work}
\label{conclusion}
In this work, we introduced an approach for 5G-specific command and control communication, 5G Puppeteer.
Several problems, the usable interfaces, procedures and messages, the attack success, the path selection and routing, as well as performance and stealthiness were discussed and methods to implement and deploy the approach were described.
We found that 5G Puppeteer is able to provide command and control capabilities via genuine interfaces, procedures, and messages.
Furthermore, the options to choose messages, paths, and embedding locations are extensive.
While only example solutions were presented, the extensive diversity suggests that effective countermeasures are complex to implement and that 5G Puppeteer can most likely be deployed in an undetectable fashion.

From the provided discussion, three suggestions can be derived in addition to the countermeasures discussed in section \ref{sec:countermeasures}.
First, the \ac{amf} as central component for most attacks plays a crucial role in overall system security.
Thus, the \ac{amf} should be secured and monitored most rigorously, e.g., with regular audits and host-based intrusion detection.
Second, transient messages allow skipping some of the most secured and hard to compromise components as discussed for the \ac{aka} sub-procedure.
Therefore, when it comes to network monitoring, an emphasis should be on information that is forwarded unmodified.
Third, information bridging can be prevented by avoiding transient message parameters in most vulnerable attack paths.
An example in the context of the previously discussed \ac{aka}-based \ac{udm} key extraction is to include the \ac{seaf}, \ac{ausf}, and \ac{udm} in the generation of RAND and AUTN.
With this, transient messaging is not possible for these parameters.
Instead, an attacker would need to compromise each component on the attack path.
However, this would require a change in the standardization of the \ac{aka} procedure.

\balance

\bibliographystyle{ACM-Reference-Format}
\bibliography{bibfile.bib}
\end{document}